\documentclass[aps,amssymb,amsmath,pra,reprint,noshowpacs]{revtex4-1}
\usepackage{times}
\usepackage{amssymb,amsmath,graphicx}
\usepackage[usenames]{color}
\usepackage{bbold}

\usepackage[T1]{fontenc}
\usepackage[utf8]{inputenc}
\usepackage[english]{babel}

\usepackage{textcomp}

\newcommand{\figwidth}{0.9\columnwidth} 
\newcommand{\commentOut}[1]{}
\usepackage{scalefnt}   

\newcommand{\affil}{Photonics Laboratory, ETH Zürich, CH-8093 Zürich, Switzerland}

\begin{document}
\scalefont{1.05}
\title{Controlling the net charge on a nanoparticle optically levitated in vacuum}
\author{Martin Frimmer}
\affiliation{\affil}
\homepage{http://www.photonics.ethz.ch}
\author{Karol Luszcz}
\affiliation{\affil}
\author{Sandra Ferreiro}
\affiliation{\affil}
\author{Vijay Jain}
\affiliation{\affil}
\author{Erik Hebestreit}
\affiliation{\affil}
\author{Lukas Novotny}
\affiliation{\affil}


\begin{abstract}
 Optically levitated nanoparticles in vacuum are a promising model system to test physics beyond our current understanding of quantum mechanics. Such experimental tests require extreme control over the dephasing of the levitated particle's motion. If the nanoparticle carries a finite net charge, it experiences a random Coulomb force due to fluctuating electric fields. This dephasing mechanism can be fully excluded by discharging the levitated particle. Here, we present a simple and reliable technique to control the charge on an optically levitated nanoparticle in vacuum. Our method is based on the generation of charges in an electric discharge and does not require additional optics or mechanics close to the optical trap.
\end{abstract}
\date\today

\maketitle

\paragraph{Introduction.}
An endeavor at the forefront of contemporary physics is to test the limits of quantum mechanics, in particular regarding the decoherence of quantum states of matter~\cite{Hornberger2003,Romero-Isart2011,Pikovski2012,Bassi2013,Pino2016}. In this context, a highly promising test-bed are optomechanical systems, in which the motion of a mechanical degree of freedom is controlled via the forces exerted by light fields~\cite{Aspelmeyer2014}. Great strides have been made towards bringing such mechanical oscillators from the classical Newtonian regime to the quantum ground state of motion~\cite{Teufel2011,Chan2011}. Mechanically suspended micro- and nanomechanical systems suffer from inevitable loss mechanisms due to their clamped nature. In contrast, optically levitated particles at sufficiently low pressures are potentially coupled to their surrounding only via the radiation field, promising excellent control over the decoherence of the system~\cite{Chang2010,Romero-Isart2011a}. Remarkable progress has been made to optically cool the center-of-mass motion of an optically levitated nanoparticle, putting ground-state cooling firmly within reach~\cite{Li2011,Gieseler2012,Millen2015,Fonseca2016,Jain2016,Vovrosh2016}. While at moderate pressures, the coupling to the surrounding gas is the dominant interaction dephasing the levitated particle's motion, at sufficiently low pressures the radiation pressure shot noise of the light field trapping the particle emerges as another significant damping mechanism~\cite{Jain2016}. This optical dephasing mechanism, a consequence of the quantized nature of the light field trapping the particle, might be controlled using engineered quantum states of light~\cite{Clark2017}.
While dephasing mechanisms due to optical and thermal interactions have been at the center of attention~\cite{Millen2014}, the influence of fluctuating electrostatic forces acting on optically levitated nanoparticles have remained {experimentally} unaddressed thus far. These forces arise from fluctuating fields in the environment through the Coulomb interaction, if the levitated particle carries a net charge~{\cite{Ugolini2008,Turchette2000}}. On the one hand, controlling the net charge on the particle, and in particular discharging it to zero net charge, could fully exclude any dephasing due to such fluctuating fields. On the other hand, being able to adjust the net charge on the particle to a known finite value provides a reliable handle for exerting a known force by applying an electric field~{\cite{Ashkin1980,Ranjit2015,Ranjit2016}}.
For optically levitated micrometer-sized dielectric spheres, two methods for changing the net charge have been described. First, the emission of photoelectrons in a multi-photon process was reported for microspheres levitated in laser fields at visible frequencies~\cite{Ashkin1976}. This effect can be used only for discharging initially negatively charged particles and does not allow for reversible charging and discharging. Furthermore, controlled charging and discharging of optically levitated microspheres using a single-photon photoelectric effect has been demonstrated~\cite{Ashkin1980,Moore2014,Rider2016}.
{This approach requires optical access to the particle at ultraviolet wavelengths to remove electrons. Unfortunately, such wavelengths are incompatible with the typical infrared optics used for trapping.}
Furthermore, the limited cross section for photoelectron emission for sub-micron sized particles may be a reason why controlling the net charge on an optically levitated nanoparticle has remained elusive to date.

In this paper, we report a method to measure and control the net charge on an optically levitated nanoparticle in vacuum. Our approach stands out due to its simplicity, requiring no additional optical or mechanical components beyond those required for trapping a particle and detecting its motion. The reported technique harnesses an electric discharge at low vacuum and enables us to control the net charge on the levitated particle with single elementary-charge precision, including zero net charge, before bringing it to high-vacuum without changing its charge state.

\paragraph{Experimental setup.}
\begin{figure}
\includegraphics[width=\figwidth]{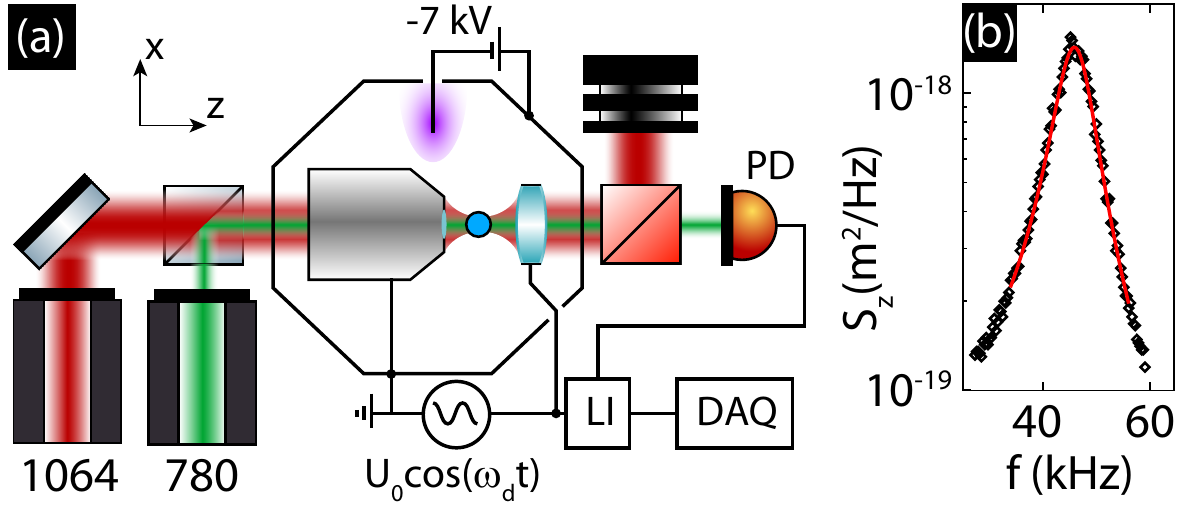}
\caption{(a)~Experimental setup. A silica nanoparticle is trapped in a focused laser beam (1064\,nm). A second laser (780\,nm) is used to measure the position of the particle. The metal housings of the microscope objective and of the lens collecting the scattered light form a capacitor to generate a low-frequency electric field at the particle position. The signal of the function generator driving the capacitor serves as a reference to a lock-in amplifier (LI) used to demodulate the photodetector (PD) signal. The LI-output is recorded by a data-acquisition card (DAQ). A wire reaching into the vacuum chamber is connected to a high-voltage source to ionize residual gas molecules and provide charges to the particle.
(b)~Power spectral density of particle position along the optical axis at a pressure of 10\,mbar.
}
\label{fig:setup}
\end{figure}
Our experimental setup is shown in Fig.~\ref{fig:setup}(a). A laser beam (1064\,nm, 50\,mW) is focused by an objective (100$\times$, NA\,0.9) to a diffraction limited spot forming an optical dipole trap. We trap silica particles with a nominal diameter of 136\,nm and mass $m=3\times10^{-18}$\,kg. To load the trap, we spray small droplets of a solution of particles suspended in isopropanol into the focal region. To infer the particle position, we use a separate measurement beam (780\,nm, 3\,mW), coaligned with the trapping beam. Both beams, together with the scattering from the particle, are recollimated after the trap by a collection lens. After dumping the trapping beam, the measurement beam is directed to a photodetector, where the interference of the measurement beam with the scattering from the particle provides a signal proportional to the particle's position along the optical axis, which we choose to be the $z$-axis. To lowest order in displacement, the particle's center-of-mass motion resembles a harmonic oscillator. Figure~\ref{fig:setup}(b) shows the power spectrum of the detector signal at a pressure of 10\,mbar.  The spectrum of the thermally driven particle is described well by a Lorentzian function [solid line in Fig.~\ref{fig:setup}(b)], whose {angular} center frequency $\Omega_z$ is given by the stiffness of the optical trap. The width of the Lorentzian is set by the damping rate $\gamma$ due to the viscous damping by the surrounding gas. Finally, the area under the Lorentzian power spectrum by definition equals the variance $\langle z^2\rangle$ of the particle's position, which has to fulfill the equipartition theorem $m\Omega_z^2\langle z^2\rangle=k_BT$, where $k_B$ is Boltzmann's constant and $T$ is room temperature. Making use of the equipartition theorem, we can convert the detector output voltage into a position in meters.
\begin{figure}
\includegraphics[width=\figwidth]{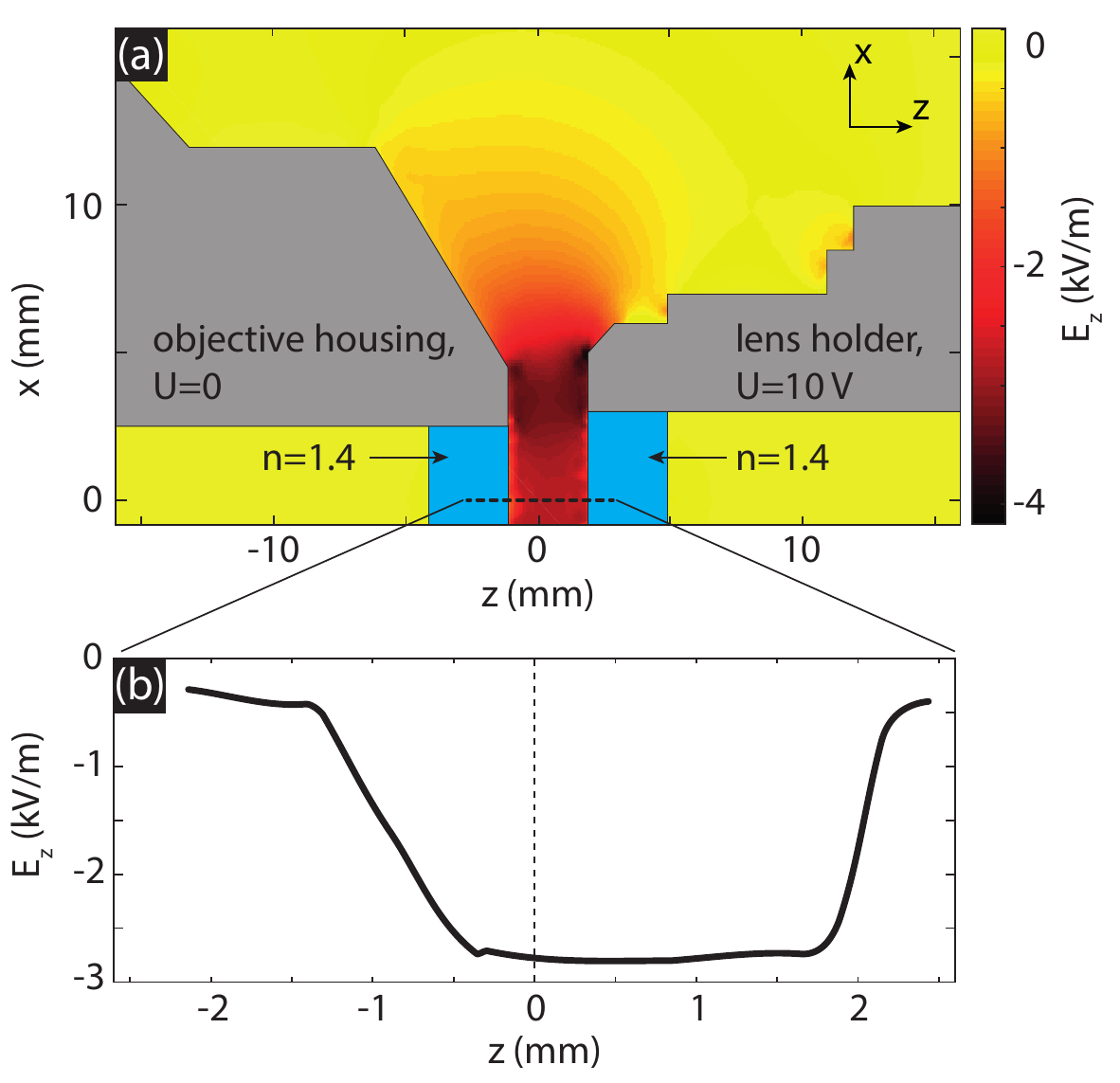}
\caption{(a)~False color plot of the simulated electric field strength $E_z$ generated by a potential of 10\,V applied to the metallic collection lens holder with the metallic objective housing grounded. The refractive index of the optics is $n=1.4$. (b)~Cross section along dashed line in (a), showing field strength $E_z$ along the optical axis. The focal plane is at $z=0$.
}
\label{fig:fieldSim}
\end{figure}

\begin{figure*}
\includegraphics[width=1.8\columnwidth]{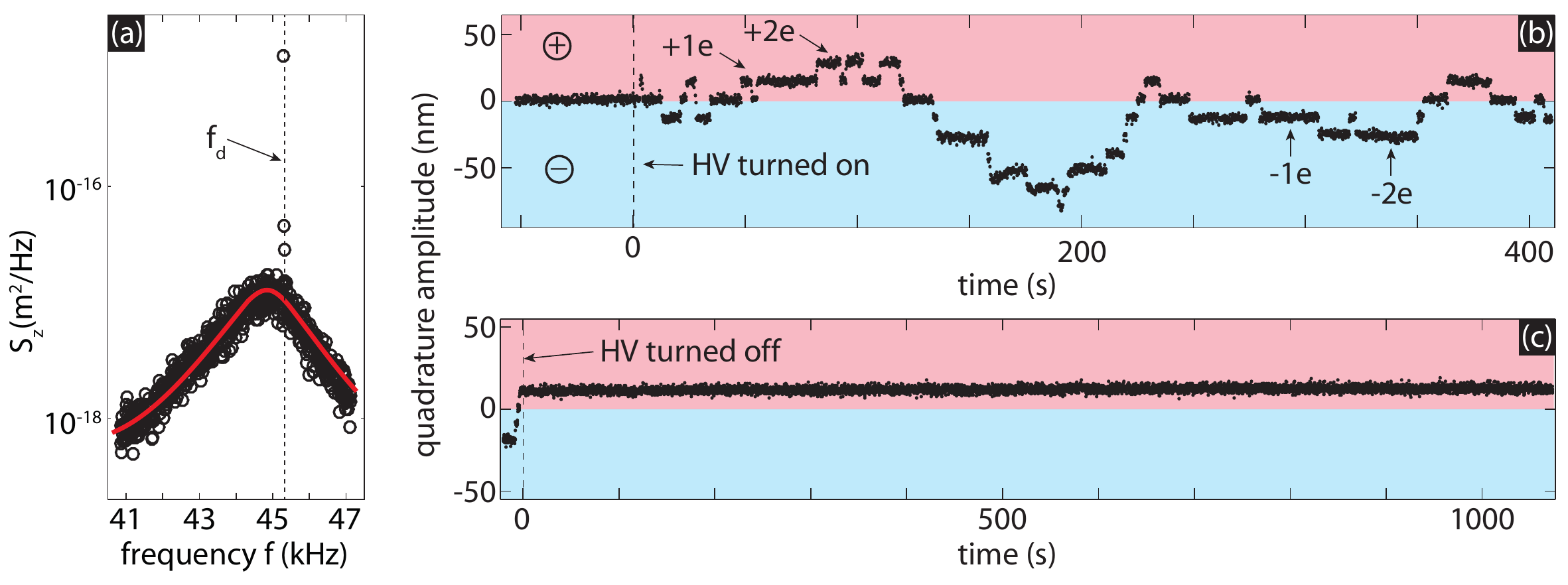}
\caption{(a)~{Power spectral density $S_z$ of the motion along the optical axis of a charge-carrying particle at a pressure of 1.9\,mbar in the presence of a drive tone at $f_d=\omega_d/(2\pi)$ applied to the capacitor.} The solid line is a Lorentzian function fit to the data. (b)~Quadrature component of particle oscillation in response to a driving voltage $U_0=10$\,V, demodulated in a bandwidth of 7\,Hz. The high-voltage (HV) discharge is turned on at $t=0$. The oscillation amplitude changes in discrete steps while the high-voltage is on. (c)~Preparation of charge state. The high-voltage is turned off at $t=0$, while the particle carries a net charge of $1e$. The charge stays constant over the remainder of the measurement.}
\label{fig:AmpPhase}
\end{figure*}

To detect the net charge on the particle, we drive the particle motion with an electric field. To this end, we ground the metal housing of the microscope objective and apply a voltage $U(t)=U_0\cos(\omega_d t)$ to the metal holder of the collection lens, thereby forming a capacitor, as depicted in Fig.~\ref{fig:setup}(a). This method of applying an electric field does not require any additional mechanical components in the setup and therefore does not obstruct access to the particle. Figure~\ref{fig:fieldSim}(a) shows a false-color plot of the electric field strength $E_z$ simulated in a plane containing the optical axis, when a potential of 10\,V is applied to the collection lens holder. When we apply a voltage to the capacitor at a frequency $\omega_d$ close to the particle's resonance frequency $\Omega_z$, we observe the response of the particle to the driving field as a distinct peak in the power spectrum in addition to the thermal population, as shown in Fig.~\ref{fig:AmpPhase}(a). We demodulate the detector signal at the driving frequency with a lock-in amplifier.

Our method to control the charge on the levitated nanoparticle relies on ionizing gas molecules inside the vacuum chamber~\cite{Schonenberger1990}. We use a homebuilt Cockcroft-Walton voltage multiplier to generate a DC voltage of $-7$\,kV, which we feed into the vacuum chamber to a bare wire of about 5\,cm length. This wire is about 5\,cm away from the optical trap. The grounded vacuum chamber serves as the counter electrode.

\paragraph{Results and discussion.} In Fig.~\ref{fig:AmpPhase}(b), we show a typical time trace of the demodulated detector signal in quadrature with the drive, expressed as the oscillation amplitude of the particle, at a drive voltage $U_0=10$\,V\@. The phase of the reference signal has been adjusted for the in-phase component of the signal to vanish, despite a small detuning of the drive frequency $\omega_d$ from the particle's resonance $\Omega_z$. The high-voltage is switched on at time $t=0$ in Fig.~\ref{fig:AmpPhase}(b). As soon as the high-voltage is on, the signal changes in discrete steps, which we interpret as a signature of single elementary-charge transfer to and from the nanoparticle.
The demodulated signal can assume both positive and negative values, since the response to the driving field flips phase by $\pi$ when the polarity of the charge changes. Knowing the transfer functions of our electronics, we can associate states with positive (negative) net charge to positive (negative) signal amplitudes. In Fig.~\ref{fig:AmpPhase}(b), we indicate regions of positive (negative) net charge with red (blue) shading.

To demonstrate the stability of the charge state in the absence of the discharge, we show the demodulated detector signal for an experiment where the high-voltage is switched off at time $t=0$, while the particle carries the net charge $1e$, shown in Fig.~\ref{fig:AmpPhase}(c). Clearly, the charge state of the particle remains constant over the remainder of the measurement. Importantly, we have never observed the charge state of the particle change while the high-voltage source is turned off, even over a period of several days.
Furthermore, we have discharged particles at a pressure around 1\,mbar and subsequently brought them to pressures in the range of $10^{-5}$\,mbar.
At this low pressure, we have never observed any measurable response of the particle to a driving field.
This observation means that the particle's charge state is preserved when the pressure is reduced. Accordingly, by simply turning off the high-voltage source when the particle carries the desired number of charges, the levitated particle can be brought to any desired net-charge state from zero to a few elementary charges, before reducing the pressure to the high-vacuum regime, where levitated-optomechanics experiments are typically conducted~\cite{Jain2016}. For completeness, we note that our particles right after trapping typically carry a charge of several ten positive elementary charges. After turning on the high-voltage, within a few seconds, the net charge drops to values of only a few elementary charges.
{As a word of caution, we note that using an ion pressure gauge, if not properly shielded, can cause changes in the charge state of the particle.}
Regarding the mechanism of charge transfer to the levitated particle, we suggest that the high-voltage ionizes gas molecules inside our vacuum chamber. Indeed, we observe the typical purple emission of a glow discharge in close proximity of the high-voltage wire forming the cathode. The trapped particle is located in the Faraday dark space of the plasma, where the net charge is close to neutral~\cite{Lieberman2005}. {The fact that the net charge on the particle can change from positive to negative clearly shows that the electric discharge indeed provides charges of both polarities to the particle. These charges can be both electrons and ionized air molecules generated in collisions with charges accelerated by the high-voltage~\cite{Lieberman2005}.} We conjecture that the attraction of charges of opposite polarity further helps to keep the net charge on the particle within less than typically ten elementary charges. {Furthermore, we note that the long-time average of the charge on the nanoparticle is $-1.7e$ in the measurement shown in Fig.~\ref{fig:AmpPhase}(b). We speculate that this fact could be a signature of a finite net charge density at the particle position, expected for asymmetric electrode geometries~\cite{Bogdanov2015}.
We also point out that the rate at which the charge jumps occur peaks at pressures around $0.1$ to $1$~mbar, an observation that could indicate a maximum in the density of ionized species present in the chamber.}

We now provide a further cross-check of our claim that the plateaus observed in Fig.~\ref{fig:AmpPhase}(b) are indeed due to single elementary-charge transfer. From the power spectrum of the thermally driven particle [red line in Fig.~\ref{fig:AmpPhase}(a)], we know the damping rate $\gamma$ and the eigenfrequency $\Omega_z$, defining the transfer function of the harmonic oscillator. Accordingly, from the amplitude measured at the single-charge plateau, and assuming the currently accepted value for the elementary charge $e$, we can deduce the field strength $E_z$ acting on the particle along the optical axis. We obtain a value of $E_z=-2.72$\,kV/m, which is in good agreement with the simulation of the field at the trap position shown in Fig.~\ref{fig:fieldSim}(b), yielding a value of $E_z^\mathrm{sim}=-2.77$\,kV/m.

Let us turn to the analysis of the sensitivity of our charge measurement. Our technique is based on sensing the Coulomb force acting on the particle.
We expect the sensitivity of our charge measurement to be limited by thermal effects, since at the driving frequency $\omega_d$ the thermal population of the particle largely exceeds any other noise present in the system, as shown in Fig.~\ref{fig:AmpPhase}(a). In our case, the thermal population of the oscillator in the measurement bandwidth corresponds to a standard deviation of the position $\sqrt{\langle z_\text{thermal}^2\rangle}=3.4\,\text{nm}$. This value is in good agreement with the residual measured amplitude $\bar z_\text{0}=2.5\,\text{nm}$ at the zero-charge state in Fig.~\ref{fig:AmpPhase}(b). {Accordingly, our charge sensor is operating at the thermal sensitivity limit, where the noise-equivalent charge $q_\text{ne}$ is determined by the strength of the thermal force (set by the temperature $T$ and the damping rate $\gamma$) within the measurement bandwidth $\Delta f$, leading to $\langle q_\text{ne}^2\rangle=4k_BTm\gamma\Delta f/E^2$, with $E$ the applied electric field~\cite{Braginsky1985}.} At the operating conditions discussed in Fig.~\ref{fig:AmpPhase}, the noise-equivalent charge is $q_\text{ne} = 0.15e$, which agrees with the experimental accuracy observed in Fig.~\ref{fig:AmpPhase}(b).
We note that the charge sensitivity can be boosted by operating at lower gas pressures and lower gas temperatures.

\paragraph{Conclusion.} In conclusion, we have presented a method to measure and control the charge on an optically levitated nanoparticle in vacuum. Importantly, being sensitive to the charge-to-mass ratio, our method is equally suitable to measure the mass of optically levitated objects after adjusting their charge. This ability may prove useful in the characterization of more complex levitated objects, fabricated with limited homogeneity~\cite{Neukirch2015,Rahman2017}. Our approach is well suited to be implemented in more elaborate trapping setups, especially those involving optical resonators with the goal of ground-state cooling~\cite{Kiesel2013,Fonseca2016}, where access to the optical trap is usually heavily restricted. Furthermore, the possibility to control the charge on a levitated nanoparticle generates new opportunities for feedback cooling by applying a direct damping force~\cite{Poggio2007}, for force sensing, including probing surface interactions at microscopic length scales~\cite{Stern1988,Terris1989,Bressi2002,Schonenberger1990}, and for exploiting novel non-linear sensing principles~\cite{Leuch2016}. {Finally, our work suggests that a levitated nanoparticle could be used as a sensitive probe to investigate
non-equilibrium physics on a microscopic scale inside a plasma~\cite{Gieseler2014,Schablinski2015}.}

\begin{acknowledgments}
This research was supported by ERC-QMES (Grant No.~338763) and the NCCR-QSIT program (Grant No.~51NF40-160591).
\end{acknowledgments}

\bibliography{Frimmer_Literature}

\end{document}